# Dynamics of a small planetoid in Newtonian gravity field of Lagrangian configuration of three primaries

**Sergey Ershkov***, Affiliation[1]: Sternberg Astronomical Institute,

M.V. Lomonosov's Moscow State University, 13 Universitetskij prospect, Moscow 119992, Russia, Affiliation[2]: Plekhanov Russian University of Economics,

Scopus number 60030998, e-mail: sergej-ershkov@yandex.ru,

**Dmytro Leshchenko**, Odessa State Academy of Civil Engineering and Architecture, e-mail: leshchenko_d@ukr.net,

**Alla Rachinskaya**, Odessa I. I. Mechnikov National University,

e-mail: rachinskaya@onu.edu.ua

## Abstract

Novel method for semi-analytical solving of equations of a *trapped dynamics* for a planetoid $m_4$ close to the plane of mutual motion of main bodies around each other (in case of a special type of Bi-Elliptic Restricted 4-Bodies Problem) is presented. We consider here three primaries $m_1$, $m_2$, $m_3$ orbiting around their center of mass on *elliptic* orbits which are permanently forming Lagrangian configuration of an equilateral triangle. Our aim is to obtain appoximate coordinates of quasi planar trajectory of the infinitesimal planetoid $m_4$, when the primaries have masses equal to 1/3. Results are as follows: 1) equations for coordinates $\{\bar{x},\ \bar{y}\}$ are described by system of coupled 2-nd order ordinary differential equations with respect to true anomaly $f$, 2) expression for $\bar{z}$ stems from solving second order *Riccati* ordinary differential equation that determines the *quasi-periodical* oscillations of planetoid $m_4$ not far from invariant plane $\{\bar{x}, \bar{y}, 0\}$.

## 1. Introduction.

It is well-established fact in Celestial Mechanics that equations of the restricted 3-bodies problem [1-3], hereafter refereed as R3BP, is non-integrable (where we consider orbiting of infinitesimal planetoid gravitationally influenced by duet of 2 primaries $M_{Sun}$ and $m_{planet}$, both orbiting in Kepler-type dynamics, $m_{planet}$ is assumed to be less than $M_{sun}$; we should note that the hypothesis that one of the primaries has a bigger mass than the other is basically extraneous to the problem: it enters the game later, when reseacher's attention is focused on practical applications). This is a classical well known problem: two primaries revolve on Keplerian orbits around the barycentre, and a third body, named planetoid, moves under the gravitational attraction of the primaries, without affecting their motion. If the motion of the primaries is circular then one obtains the classical R3BP. It is worth to note a valuable contributions of famous participants of Celestial Mehanics community regarding analytical methods and the obtained results in R3BP [1-7], stability of solutions [8-10], input of such findings to investigation of influence of tidal phenomena on dynamics of fluidic envelope of fluid-type planets and their rotational dynamics along with their satellites [11-12], variety of partial formulations and applications in Celestial Mechanics [13-14] (including of those of non-gravitational nature [15]), investigations of stability of the solutions [16-20] (including of those in the vicinity of libration points [21-24]), and partial applying the aforedescribed findings in investigations of various non-linear perturbing effects and phenomena [25-26], e.g. in investigations of the escape and collision dynamics as well [27]. Especially, let us mention a case of elliptic restricted 3-bodies problem [4-5] (ER3BP, where primaries are orbiting not far from their center of masses on *elliptic* orbits) and, also, of BiER4BP, Bi-Elliptic Restricted 4-Bodies Problem [6-7]. The problem investigated here may be enunciated as follows: let three primaries move on homographic elliptic orbits in a triangular configuration; study the dynamics of a fourth body, subject to the attraction of the primaries without affecting their orbits. The existence of homographic triangular motions for the three body problem is known since Lagrange's time. The idea exploited in the paper is: use the triangular homographic solutions as a basis, in the same sense that homographic solutions of a two body system are the basis for the restricted problem; then study the motion of a planetoid.

Following by the ansatz in [6] for BiER4BP, we will consider here 3 primaries $m_1$, $m_2$, $m_3$ which are supposed in our research to be moving not far from center of masses in *elliptic* orbital mutual motions on homographic triangular orbits, Fig.1 (in general case, homographic solutions are not restricted in size):

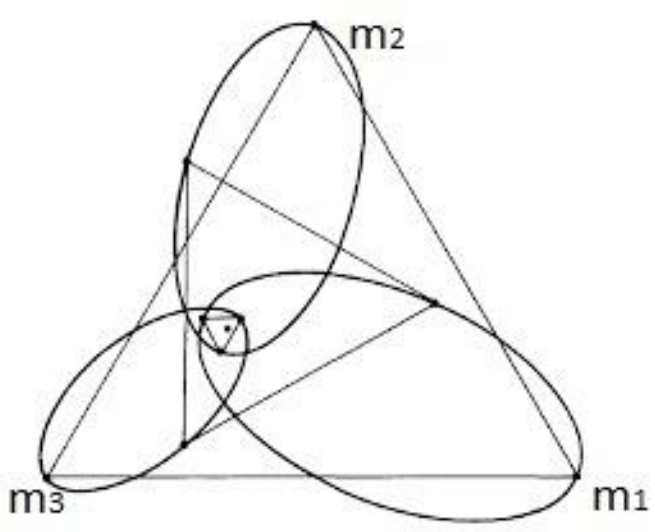

Fig.1. Lagrange solution (equilateral triangle) to the three-body problem.

Let us note that we schematically imagine on Fig.1 the motions of three primaries which are forming the equilateral triangle in their mutual motion (such the Lagrange solution is able to be presented of non-equal masses located in the vertices of above mentioned triangle, in general case).

It is worthnoting that motivation, method of semi-analytical investigation of the problem and presented final results of the current research obviously differ from those presented in [18] (and so such the results should be considered as novel). In paper [18] the case of *trapped* orbiting of infinitesimal mass, which is moving *exclusively* in the vicinity of the smallest from triplet of primaries in alternative variant of BiER4BP (with respect to presented here) has been studied. In the present paper we extend our study to the case when planetoid is freely moving between all three primaries. Besides, in [18] all three primaries are orbiting around center of masses in elliptic orbits with *hierarchical* assumption regarding masses of bodies where masses of two smallest primaries (distant planets) much less than the mass of main primary (Sun); the aforementioned formulation of the problem [18] quite differs from formulation of the problem, presented in the current research: we explore here motion or quasi planar trajectory of the infinitesimal planetoid $m_4$, when all primaries have masses equal to 1/3 (with respect to the combined mass of all 3 primaries). Finally, let us outline that various approximations considered along with the mathematical development, presented in the current research, must be specified since the beginning of the paper because they restrict the range of applicability of the presented approach. Among these assumptions are equal masses for the primaries, low (and equal) eccentricities of the trajectories of all the primaries (homographic orbits are by definition similar, hence the eccentricities of the primaries need be equal), and small values for the coordinate $\bar{z}$, in *pulsating* system of rotating coordinates $\{\bar{x}, \bar{y}, \bar{z}\}$ (this assumption will be presented and discussed in the next section). The introduction of a (non uniformly) rotating and pulsating reference system in order to describe elliptic homographic orbits for the three body problem is a standard tool (see for instance [2]).

## 2. Description of the model, equations of motion.

Let us present in the current investigation a novel method for resolving equations of a *quasi planar orbiting* of infinitesimal planetoid $m_4$ close to the plane of mutual orbitings of the main primaries around each other (in case of a special type of BiER4BP) is presented with illuminating the results by numerical findings.

We consider three primaries $m_1$, $m_2$, $m_3$ which are orbiting not far from their center of mass on *elliptic* orbits with permanent geometrically forming of an equilateral triangle for them (Fig.1). Our aim for constructing the aforementioned *quasi planar motion* is the obtaining coordinates of infinitesimal planetoid $m_4$ which maintain its orbit located close to plane of orbiting primaries $m_1$, $m_2$, $m_3$ (of equal mass $m$, for the sake of simplicity). It is well-known fact that the locations of the main bodies are variable in ER3BP for the reason of their mutual Kepler-elliptic motion, so we can conclude that relative distance $\rho$ between them is also variable at such the elliptic motion [2]

$$\rho = \frac{a(1-e^2)}{1+e\cos f} \qquad (1)$$

where $a$ is semimajor axis of *elliptic* orbit of chosen primary in its motion around barycenter of all primaries ($e$ is eccentricity), $f$ is the true anomaly of chosen primary (planet) in its elliptic motion around the common center of masses of all three primaries (we should note that the eccentricity $e$ and the true anomaly $f$ are the same for all primaries; the reference to a chosen primary is not appropriate, except for the semimajor axis $a$); besides, here the distance-scaling results in $a = 1$. In addition, angular motion is given by

$$\frac{df}{dt} = \left(\frac{GM}{a^3(1-e^2)^3}\right)^{\frac{1}{2}}(1+e\cos f)^2 \qquad (2)$$

where $G$ is the constant of Newton gravity law (in units of International System of physical measurment SI, a coherent system of units of measurement), $M$ is the sum of masses of all 3 primaries (the total mass of the system). The time-scaling results in $G = 1$ in (2). Following the approach [6], system of equatons of BiER4BP for a planetoid $m_4$ is able to be represented in *pulsating-rotating* system of Cartesian coordinates $\vec{r} = \{\bar{x}, \bar{y}, \bar{z}\}$ by using relation (1) for primaries (*with appropriate initial conditions*)

$$\begin{aligned} \ddot{\bar{x}} - 2\dot{\bar{y}} &= \frac{\partial\Omega}{\partial\bar{x}}, \\ \ddot{\bar{y}} + 2\dot{\bar{x}} &= \frac{\partial\Omega}{\partial\bar{y}}, \\ \ddot{\bar{z}} &= \frac{\partial\Omega}{\partial\bar{z}}, \end{aligned} \qquad (3)$$

$$\Omega = \frac{1}{1+e\cos f}\left[\frac{1}{2}\left(\bar{x}^2 + \bar{y}^2 - \bar{z}^2 e\cos f\right) + \frac{1}{3}\sum_{i=1}^{3}\frac{1}{\bar{r}_i}\right], \qquad (4)$$

$$\bar{r}_i^{\,2} = (\bar{x}-\bar{x}_i)^2 + (\bar{y}-\bar{y}_i)^2 + (\bar{z}-\bar{z}_i)^2, \quad i = 1, 2, 3 \qquad (5)$$

where, the dot presents the derivative with respect to the independent variable $f$ in (1); $\Omega$ is the proper real (non-vector) function; besides, $\{\bar{r}_i\}$ is a set of appropriate space distances of planetoid $m_4$ from primaries $\{m_i\}$, respectively. In such the pulsating rotating coordinate system, primaries are to be in the stationary positions; this fact allows us to use in (5) the appropriate space-scaled expressions for coordinates as presented below in (5.1) (another type of scaling was used earlier in [6]); besides, we choose $z_1 = z_2 = z_3 = 0$ in (5)

$$\begin{aligned} \bar{x}_1 &= \frac{a_1}{a}, & \bar{y}_1 &= 0 \\ \bar{x}_2 &= \frac{a_2\, e}{a(1-e^2)}, & \bar{y}_2 &= \frac{a_2}{a} \\ \bar{x}_3 &= \frac{a_3\, e}{a(1-e^2)}, & \bar{y}_3 &= -\frac{a_3}{a} \end{aligned} \tag{5.1}$$

where, $\{a_1, a_2, a_3\}$ are semimajor axes of elliptic-orbits of primaries $\{m_1, m_2, m_3\}$ around the common center of masses, accordingly. Then further, expressions (5.1) can be simplified in case of equal masses $m$ of all primaries as in case considered in [6] (in addition, we can choose in the current research in (5.1) as follows: $2a_2 = 2a_3 = a$, $a_1 = a\left(\frac{\sqrt{3}}{2} + \frac{e}{2(1-e^2)}\right)$, such the type of scaling keeps size of appropriate side $a$ of equilateral triangle equals to 1 whereas its bisector equals to $\left(\frac{\sqrt{3}}{2}\right)$ for the reason that $(x_1 - x_2) = (x_1 - x_3) = \left(\frac{\sqrt{3}}{2}\right)$); here, Eqs. (6) chosen to be differed from those presented in [6] earlier:

$$\begin{aligned} \bar{x}_1 &= \left(\frac{\sqrt{3}}{2} + \frac{e}{2(1-e^2)}\right), & \bar{y}_1 &= 0 \\ \bar{x}_2 &= \frac{e}{2(1-e^2)}, & \bar{y}_2 &= 1/2 \\ \bar{x}_3 &= \frac{e}{2(1-e^2)}, & \bar{y}_3 &= -1/2 \end{aligned} \tag{6}$$

(it is worthnoting that different sign was used by mistake in expression for $\bar{y}_3$ in Eqs. (11) in [6]). It seems to be simply a typo (all other fundamental results in [6] are obviously correct): first, we should note that if we choose $a_2 = a_3$ in Eqs. (11) in [6] this would mean that two primaries are to be at the same fixed position in a space (which is senseless); moreover, if we calculate the size of the side of equilateral triangle based on various expressions for coordinates of vertices given in Eqs. (11) in [6], we will come to conclusion that it is not equilateral triangle. Thus, we will use (here, in Eqs. (5.1)-(6), further and below) only the correct expressions as above.

## 3. Resolving the system of equations (3).

Let us aim to present Eqns. (3) (with help of (4) and (6)) in a convinient-form for further analysis. By the proper rewriting equations (3) with respect to the partial derivatives which relate to coordinates $\{\bar{x}, \bar{y}, \bar{z}\}$, this yields

$$\ddot{\bar{x}} - 2\dot{\bar{y}} = \frac{1}{1+e\cos f}\left[\bar{x} - \frac{1}{3}\left(\frac{\bar{x}-\left(\frac{\sqrt{3}}{2}+\frac{e}{2(1-e^2)}\right)}{\left((\bar{x}-\frac{\sqrt{3}}{2}-\frac{e}{2(1-e^2)})^2+\bar{y}^2+\bar{z}^2\right)^{\frac{3}{2}}} + \frac{\left(\bar{x}-\frac{e}{2(1-e^2)}\right)}{\left(\left(\bar{x}-\frac{e}{2(1-e^2)}\right)^2+(\bar{y}-\frac{1}{2})^2+\bar{z}^2\right)^{\frac{3}{2}}} + \frac{\left(\bar{x}-\frac{e}{2(1-e^2)}\right)}{\left(\left(\bar{x}-\frac{e}{2(1-e^2)}\right)^2+(\bar{y}+\frac{1}{2})^2+\bar{z}^2\right)^{\frac{3}{2}}}\right)\right],$$

$$\ddot{\bar{y}} + 2\dot{\bar{x}} = \frac{1}{1+e\cos f}\left[\bar{y} - \frac{1}{3}\left(\frac{\bar{y}}{\left((\bar{x}-\frac{\sqrt{3}}{2}-\frac{e}{2(1-e^2)})^2+\bar{y}^2+\bar{z}^2\right)^{\frac{3}{2}}} + \frac{(\bar{y}-\frac{1}{2})}{\left(\left(\bar{x}-\frac{e}{2(1-e^2)}\right)^2+(\bar{y}-\frac{1}{2})^2+\bar{z}^2\right)^{\frac{3}{2}}} + \frac{(\bar{y}+\frac{1}{2})}{\left(\left(\bar{x}-\frac{e}{2(1-e^2)}\right)^2+(\bar{y}+\frac{1}{2})^2+\bar{z}^2\right)^{\frac{3}{2}}}\right)\right], \quad (7)$$

$$\ddot{\bar{z}} = \frac{1}{1+e\cos f}\left[-\bar{z}\,e\cos f - \frac{1}{3}\left(\frac{\bar{z}}{\left((\bar{x}-\frac{\sqrt{3}}{2}-\frac{e}{2(1-e^2)})^2+\bar{y}^2+\bar{z}^2\right)^{\frac{3}{2}}} + \frac{\bar{z}}{\left(\left(\bar{x}-\frac{e}{2(1-e^2)}\right)^2+(\bar{y}-\frac{1}{2})^2+\bar{z}^2\right)^{\frac{3}{2}}} + \frac{\bar{z}}{\left(\left(\bar{x}-\frac{e}{2(1-e^2)}\right)^2+(\bar{y}+\frac{1}{2})^2+\bar{z}^2\right)^{\frac{3}{2}}}\right)\right],$$

where we use expressions (4) and (6)

$$\Omega = \frac{1}{1+e\cos f}\left[\frac{1}{2}\left(\bar{x}^2+\bar{y}^2-\bar{z}^2 e\cos f\right)+\frac{1}{3}\left(\frac{1}{\bar{r}_1}+\frac{1}{\bar{r}_2}+\frac{1}{\bar{r}_3}\right)\right],$$

$$\bar{r}_1 = \sqrt{(\bar{x}-\frac{\sqrt{3}}{2}-\frac{e}{2(1-e^2)})^2+\bar{y}^2+\bar{z}^2},\quad \bar{r}_2 = \sqrt{\left(\bar{x}-\frac{e}{2(1-e^2)}\right)^2+(\bar{y}-\frac{1}{2})^2+\bar{z}^2},\quad \bar{r}_3 = \sqrt{\left(\bar{x}-\frac{e}{2(1-e^2)}\right)^2+(\bar{y}+\frac{1}{2})^2+\bar{z}^2}\,.$$

Let us further assume that $\bar{z}$-coordinate for system (7) belongs to the sub-variety of *quasi planar orbits* for infinitesimal planetoid $m_4$, $\bar{z} \to 0$ (i.e., planetoid $m_4$ is moving in close vicinity of invariant-plane $\{\bar{x}, \bar{y}, 0\}$). Then afterwards, we should not take into account all terms less of 2nd order of smallness in (7); thus, the 3-rd Eqns. of system (7) would led us to conclusion as follows $\left(\bar{x} \neq \left(\frac{\sqrt{3}}{2}+\frac{e}{2(1-e^2)}\right), \bar{x} \neq \frac{e}{2(1-e^2)}, \bar{y} \neq \{0, \pm\frac{1}{2}\}, \bar{z} \neq 0\right)$

$$\ddot{\bar{z}} + \omega\,\bar{z} = 0\,, \tag{8}$$

$$\omega = \frac{1}{1+e\cos f}\left[e\cos f + \frac{1}{3}\left(\frac{1}{\left((\bar{x}-\frac{\sqrt{3}}{2}-\frac{e}{2(1-e^2)})^2+\bar{y}^2\right)^{\frac{3}{2}}}+\frac{1}{\left(\left(\bar{x}-\frac{e}{2(1-e^2)}\right)^2+(\bar{y}-\frac{1}{2})^2\right)^{\frac{3}{2}}}+\frac{1}{\left(\left(\bar{x}-\frac{e}{2(1-e^2)}\right)^2+(\bar{y}+\frac{1}{2})^2\right)^{\frac{3}{2}}}\right)\right] \tag{9}$$

Thus, equation (8) for resolving the coordinate $\bar{z}(f)$ stems from equations of system (7) and can be recognized as to be of class of the *Riccati* ODEs [11-12].

## 4. **Approximate solutions for Eqns. (7) which belong to *quasi planar orbits*.**

Exploiting the main assumption above (assumption of boundedness), we should further suggest solutions $\{\bar{x}, \bar{y}, \bar{z}\}$ of Eqns. (7) that are corresponding to sub-variety of quasi planar *orbits* for planetoid $m_4$ which are close to plane of orbiting of primaries $m_1$, $m_2$, $m_3$. This transforms two first Eqns. of (7) accordingly:

$$\ddot{\bar{x}} - 2\dot{\bar{y}} = \frac{1}{1+e\cos f}\left[\bar{x} - \frac{1}{3}\left(\frac{(\bar{x}-\frac{\sqrt{3}}{2})}{\left((\bar{x}-\frac{\sqrt{3}}{2}-\frac{e}{2(1-e^2)})^2+\bar{y}^2\right)^{\frac{3}{2}}} + \frac{\left(\bar{x}-\frac{e}{2(1-e^2)}\right)}{\left(\left(\bar{x}-\frac{e}{2(1-e^2)}\right)^2+(\bar{y}-\frac{1}{2})^2\right)^{\frac{3}{2}}} + \frac{\left(\bar{x}-\frac{e}{2(1-e^2)}\right)}{\left(\left(\bar{x}-\frac{e}{2(1-e^2)}\right)^2+(\bar{y}+\frac{1}{2})^2\right)^{\frac{3}{2}}}\right)\right],$$

$$\ddot{\bar{y}} + 2\dot{\bar{x}} = \frac{1}{1+e\cos f}\left[\bar{y} - \frac{1}{3}\left(\frac{\bar{y}}{\left((\bar{x}-\frac{\sqrt{3}}{2}-\frac{e}{2(1-e^2)})^2+\bar{y}^2\right)^{\frac{3}{2}}} + \frac{\bar{y}-\frac{1}{2}}{\left(\left(\bar{x}-\frac{e}{2(1-e^2)}\right)^2+(\bar{y}-\frac{1}{2})^2\right)^{\frac{3}{2}}} + \frac{\bar{y}+\frac{1}{2}}{\left(\left(\bar{x}-\frac{e}{2(1-e^2)}\right)^2+(\bar{y}+\frac{1}{2})^2\right)^{\frac{3}{2}}}\right)\right]. \tag{10}$$

Let us consider further the case $e \ll 1$ (which corresponds to the case of low (and equal) eccentricities of the trajectories of all the primaries):

$$\ddot{\bar{x}} - 2\dot{\bar{y}} = \frac{1}{1+e\cos f}\left[\bar{x} - \frac{1}{3}\left(\frac{\left(\bar{x}-\frac{\sqrt{3}}{2}-\frac{e}{2}\right)}{\left((\bar{x}-\frac{\sqrt{3}}{2})^2-e(\bar{x}-\frac{\sqrt{3}}{2})+\bar{y}^2\right)^{\frac{3}{2}}} + \frac{\left(\bar{x}-\frac{e}{2}\right)}{\left(\bar{x}^2-e\bar{x}+(\bar{y}-\frac{1}{2})^2\right)^{\frac{3}{2}}} + \frac{\left(\bar{x}-\frac{e}{2}\right)}{\left(\bar{x}^2-e\bar{x}+(\bar{y}+\frac{1}{2})^2\right)^{\frac{3}{2}}}\right)\right],$$

$$\ddot{\bar{y}} + 2\dot{\bar{x}} = \frac{1}{1+e\cos f}\left[\bar{y} - \frac{1}{3}\left(\frac{\bar{y}}{\left((\bar{x}-\frac{\sqrt{3}}{2})^2-e(\bar{x}-\frac{\sqrt{3}}{2})+\bar{y}^2\right)^{\frac{3}{2}}} + \frac{\bar{y}-\frac{1}{2}}{\left(\bar{x}^2-e\bar{x}+(\bar{y}-\frac{1}{2})^2\right)^{\frac{3}{2}}} + \frac{\bar{y}+\frac{1}{2}}{\left(\bar{x}^2-e\bar{x}+(\bar{y}+\frac{1}{2})^2\right)^{\frac{3}{2}}}\right)\right]. \tag{11}$$

## 5. **Discussions.**

We consider in the current research the dynamics of an infinitesimal planetoid governed by gravitational field of three primaries situated on the vertices of equilateral triangle. But we should make a reasonable conclusion that Eqns. (3) (presented by equations of system (7)) for a planetoid $m_4$, orbiting close to the plane of mutual motion of primaries $m_1$, $m_2$, $m_3$ (of equal mass $m$), are hard to be solved analytically for sub-class of approximated *quasi planar orbits* $\bar{z} \cong 0$ for Eqs. (7) insofar as presented herein.

Nevertheless, we have obtained from 3rd equation of (7) the *Riccati*-type Eqn. (8) for coordinate $\bar{z}$ (in case of already known solutions for coordinates $\{\bar{x}, \bar{y}\}$); so, solution $\bar{z}$ is to be oscillating quasi-periodically (close to plane $\{\bar{x}, \bar{y}, 0\}$) with appropriate variable frequency (9), depending on coordinates $\{\bar{x}, \bar{y}\}$. Then afterwards we suggest special type of approximations (10)-(11) for two first equations of system (7) with aim to obtain the *stable* numerical solutions for coordinates $\{\bar{x}, \bar{y}\}$ (at least, up to the value of true anomaly $f = 50$).

Besides, it is worth mentioning regarding the qualitative behaviour of solutions of *Riccati*-ODEs, their instability or *jumping* of the absolute values of the solutions [11-12] (in [12] *Abel* ODE of 1-st order of *Riccati*-type were investigated as well). So, we conclude that any solutions of equations (7) of motion (3) may obviously be unstable during a long period of an orbiting process of planetoid $m_4$ close to the plane of mutual motion of three primaries $m_1$, $m_2$, $m_3$ (including those ones which can be found for the libration points [21-24]). But we should especially note that the chosen assumption $\bar{z} \to 0$ (or a kind of simplification for system of equations (7) for motion of a planetoid), based on which we then obtain our semi-analytical solutions, stems from Eqn. (8) of *Riccati*-type [11-12] which is known to have quasi-periodic (and thus, bounded in magnitudes) solutions determined almost everywhere in the range of changing of true anomaly $f$.

Then hereafter, we have obtained the graphical plots of *numerical* solutions of equations (11) for coordinates $\{\bar{x}, \bar{y}\}$. We have considered the current case with assumption of low eccentricity $e << 1$.

Ending discussion, we should note that the numerical calculating for semi-analytical solutions of the 1st and 2nd Eqns. of (11) has been provided in **Appendix, A1**. The Runge-Kutta scheme of 4th order (step 0.001 proceeding from initial values) has been used for calculating the data. Graphical results of numerical calculating have been imagined on Figs. 2-6.

## 6. Conclusions.

Novel method for semi-analytical solving of equations of a *trapped dynamics* for a planetoid $m_4$ close to the plane of mutual motion of main bodies around each other (in case of a special type of Bi-Elliptic Restricted 4-Bodies Problem) is presented. Thus, the paper is concerned with the dynamics of a small body under the action of three masses (called primaries) orbiting in a triangular, elliptic homografic configuration. The problem may be considered as an extension of the classical restricted three body problem to the case of four bodies. It is restricted in the sense that the motion of the three primaries is an exact homographic solution of the three body problem, and the object of study is the motion of the fourth point. We restrict our study to the particular case of primaries with equal masses, and to orbits of the fourth body which are close to the plane of the primaries. As stated above, we have found that the dynamical equations for the fourth body may be approximated through a *Riccati* equation, thus describing a family of quasi periodic oscillations.

As a general conclusion we should also say that the problem proposed in the current research belongs to a wide class of problems of Academic interest, aimed at describing particular aspects of the dynamics of bodies under the Newtonian interaction. As such, it is a classical problem, and the method of approximation proposed in this paper is an illuminating one.

Also, the profound works should be especially mentioned which also tackled the problem under the current investigation presented herein, [28-29] (besides, it is worthnoting that non-classical effects like [30-31] have been ignored in mathematical formulation of problem here). It is also worth to note that in [29] (where BiCR3BP was investigated in a proper way, i.e. *bi-circular* restricted problem of *four* bodies) another meanings of solutions $\bar{x}_1$, $\bar{x}_2$, $\bar{x}_3$ from Eqs. (6) for positions of fixed vertices of equilateral triangle were used (in pulsating system of rotating coordinates) which, nevertheless, keep size of appropriate side of equilateral triangle equals to 1 whereas its bisector equals to $\left(\frac{\sqrt{3}}{2}\right)$ (according to the definition of equilateral triangle). Thus, such the alternative choice of coordinates for the positions of fixed vertices of equilateral triangle does not change the geometry and general dynamical properties of motion of an infinitesimal planetoid in the problem under consideration.

## Appendix (A1).

We have provided here in the current research the numerical calculating for appropriate semi-analytical solutions of the 1st and 2nd Eqns. of system (11). The Runge-Kutta scheme of 4th order (step 0.001 proceeding from initial values) has been used for calculating the data. Also, eccentricity $e$ = 0.0167 has been chosen for calculations (e.g., as in "Sun-Earth" system) for modelling the same binary mutual motions for various pairs of primaries $m_1$, $m_2$, $m_3$. Graphical results of numerical calculating have been imagined on Figs. 2-6, with the initial data presented below:

1) $\bar{x}_0 = 0.001$, $(\dot{\bar{x}})_0 = -\ 0.4$; $\bar{y}_0 = -\ 0.2$, $(\dot{\bar{y}})_0 = -\ 0.3$.

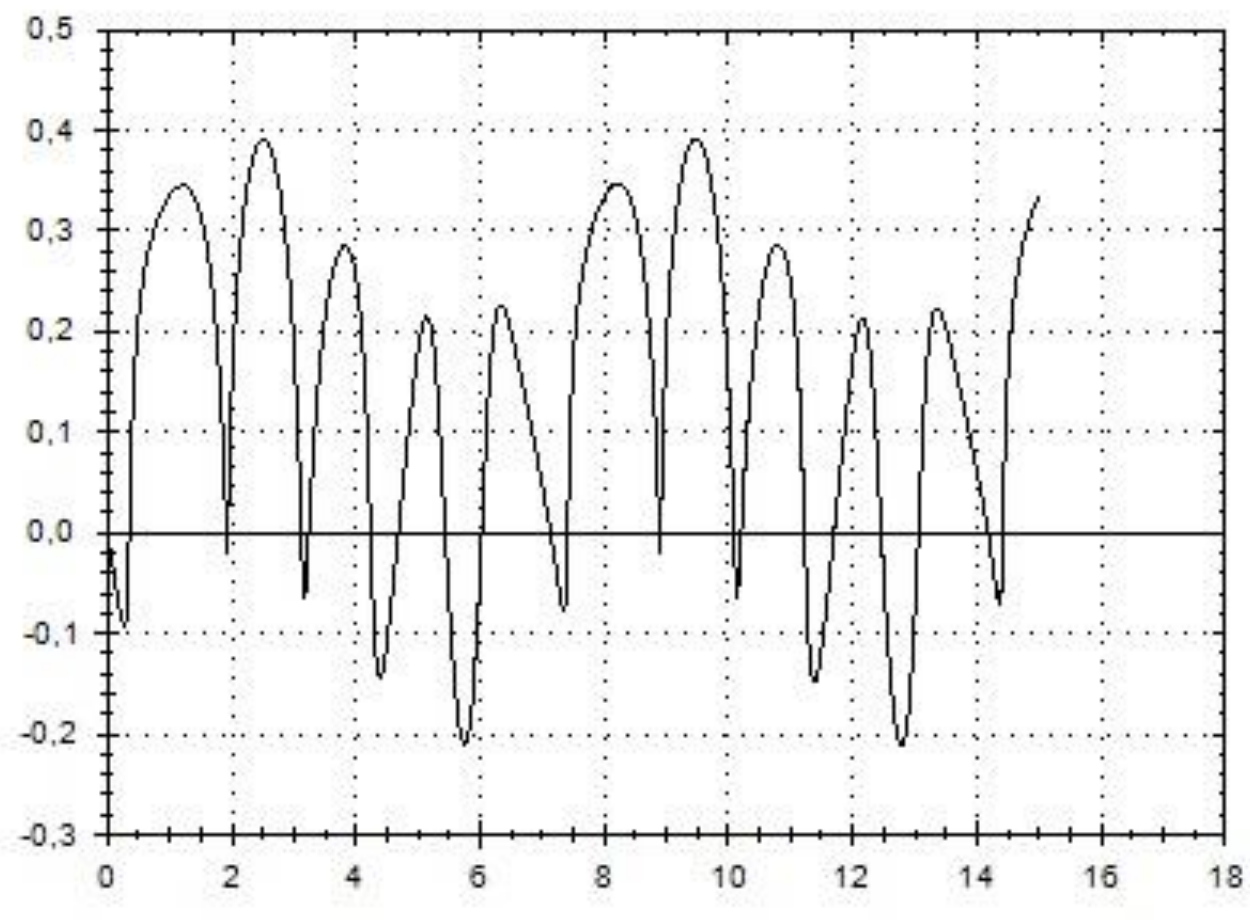

Fig.2. Numerical solution for $(\bar{x})$.

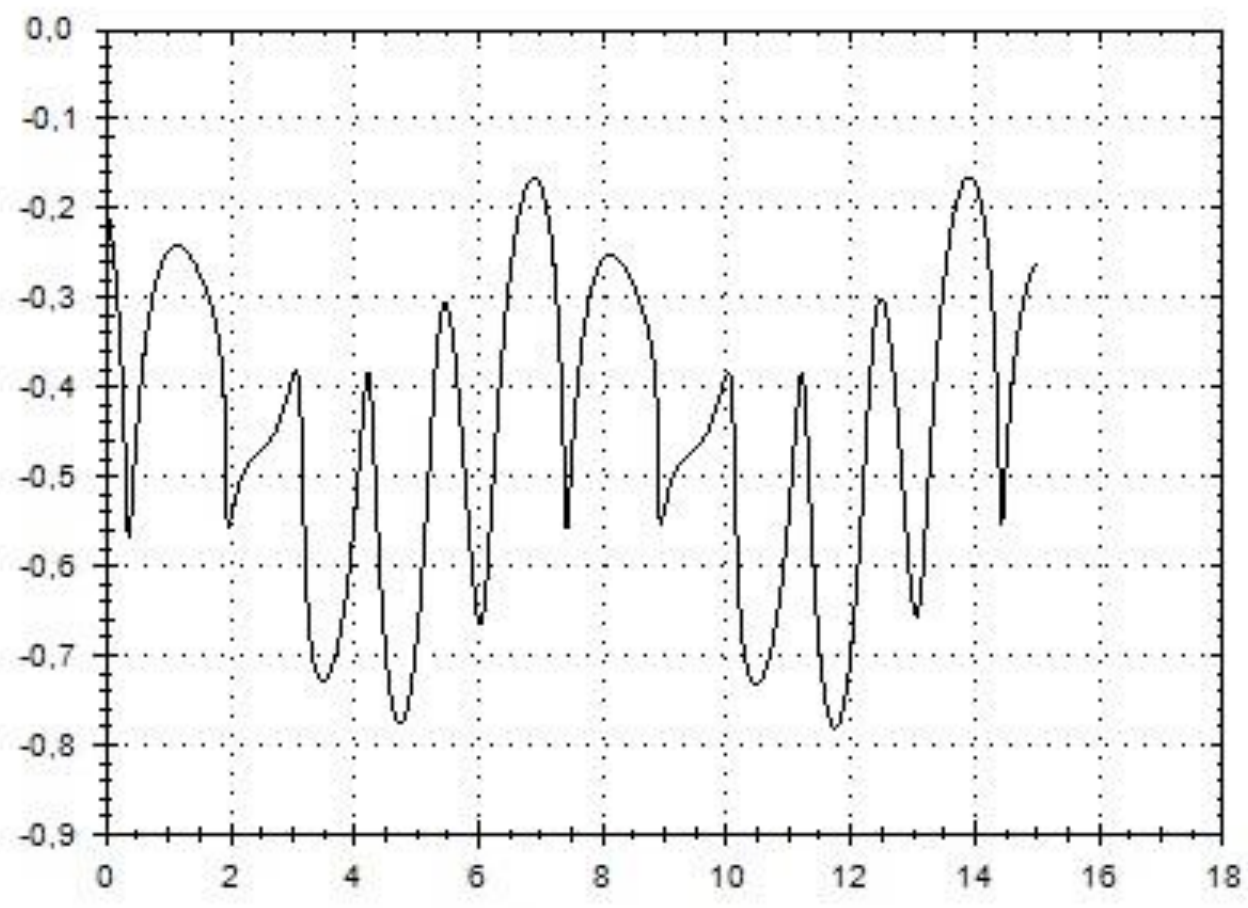

Fig.3. Numerical solution for $(\bar{y})$.

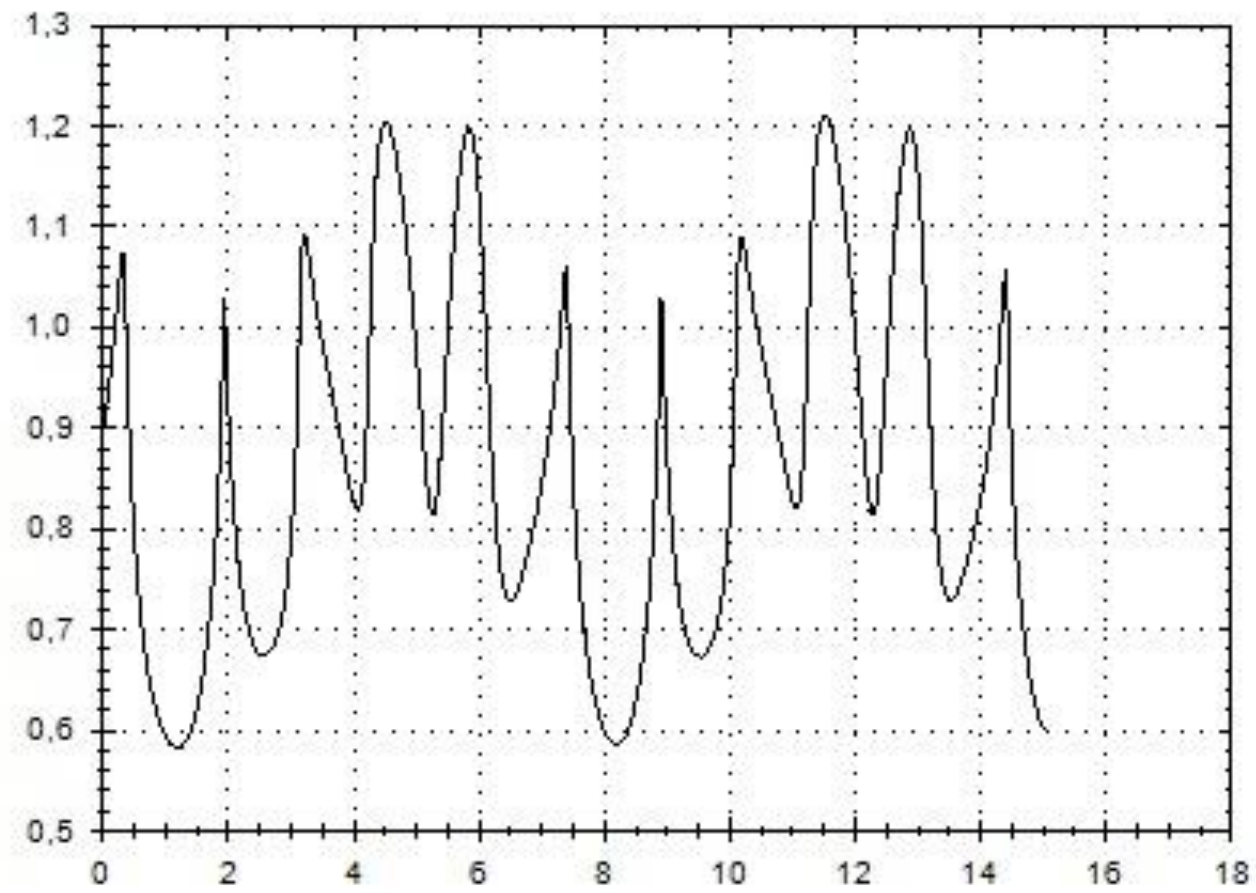

Fig.4. Numerical solution for distance $\bar{r}_1$ .

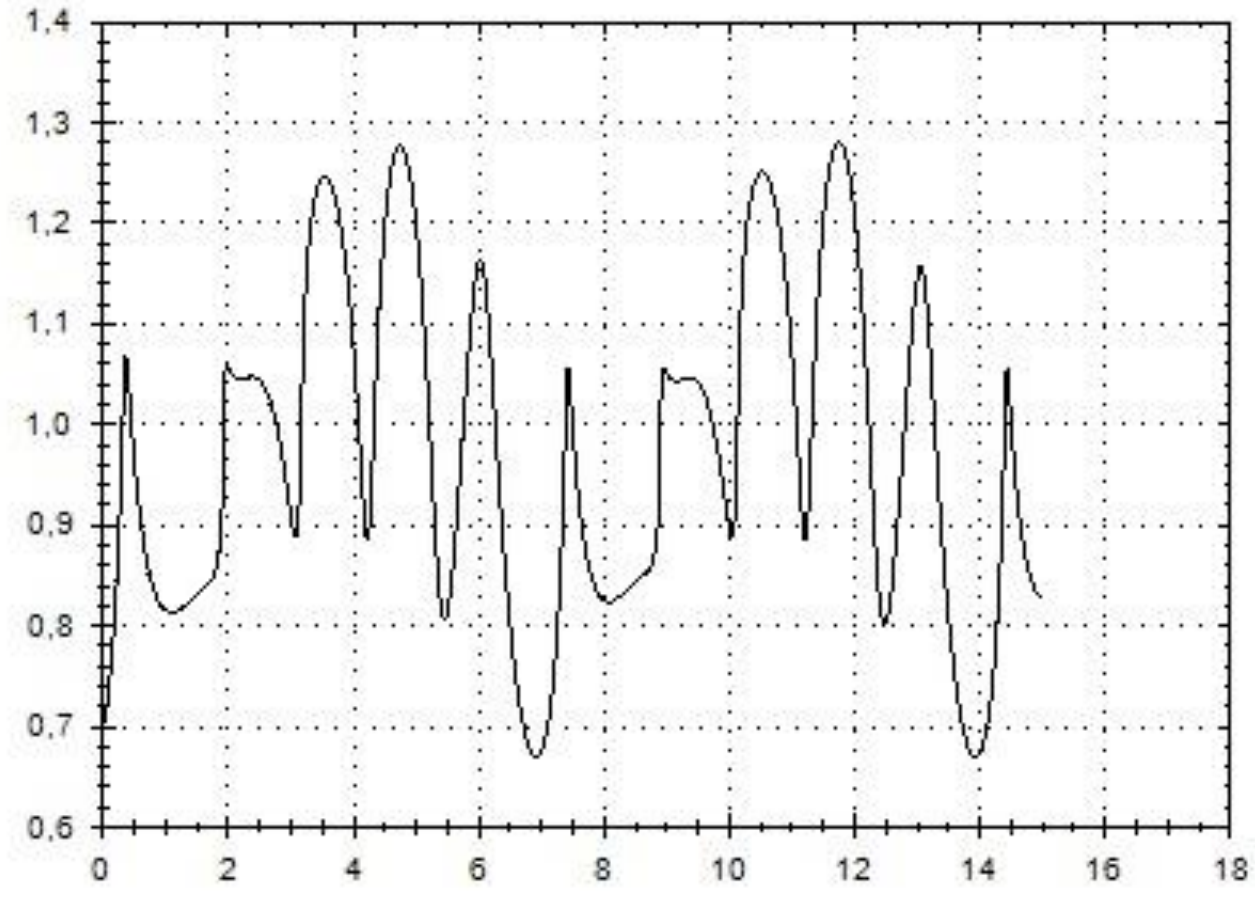

Fig.5. Numerical solution for distance $\bar{r}_2$ .

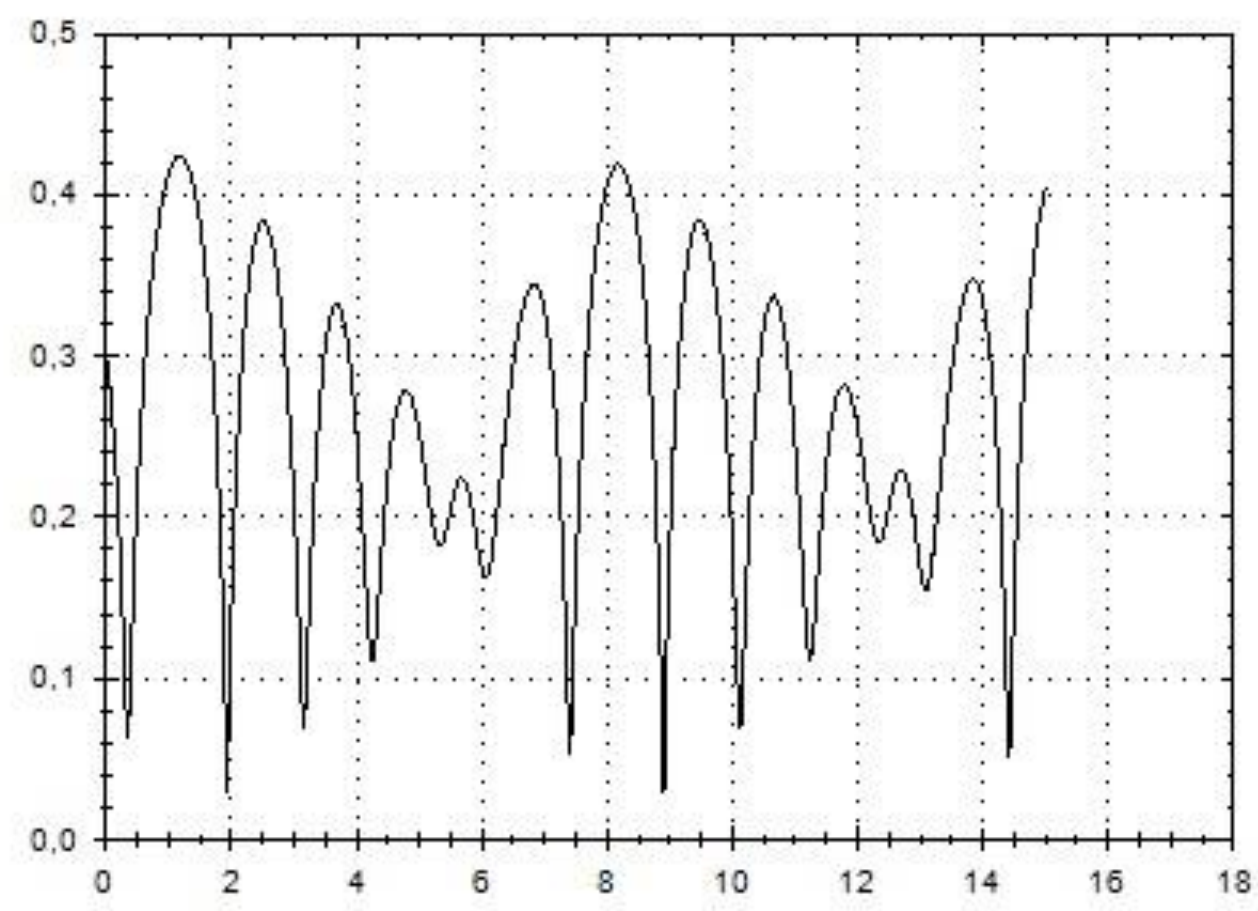

Fig.6. Numerical solution for distance $\bar{r}_3$ .

Meanwhile, it was numerically obtained for the dynamics of infinitesimal planetoid $m_4$ (see Figs.2-6) that this planetoid should be moving not far from primaries $m_1$, $m_2$, $m_3$ ($\{\bar{r}_1, \bar{r}_2, \bar{r}_3\} < 1.3$) up to the meaning of true anomaly $f \cong$ 14.5 or more than 2 full turns of first primary around the common center of masses. It is worthnoting that dynamics of components of the numerical solution is checked to be stable (at least, up to the value of true anomaly $f = 50$).

We should note that additional numerical experiments regarding solving the Eqns. (8) (with already known numerical solutions for coordinates $\{\bar{x}, \bar{y}\}$) have brought the reasonable results which can indeed be regarded as *quasi-periodical* oscillations of a planetoid in close vicinity of plane $\{\bar{x}, \bar{y}, 0\}$, see Fig.7

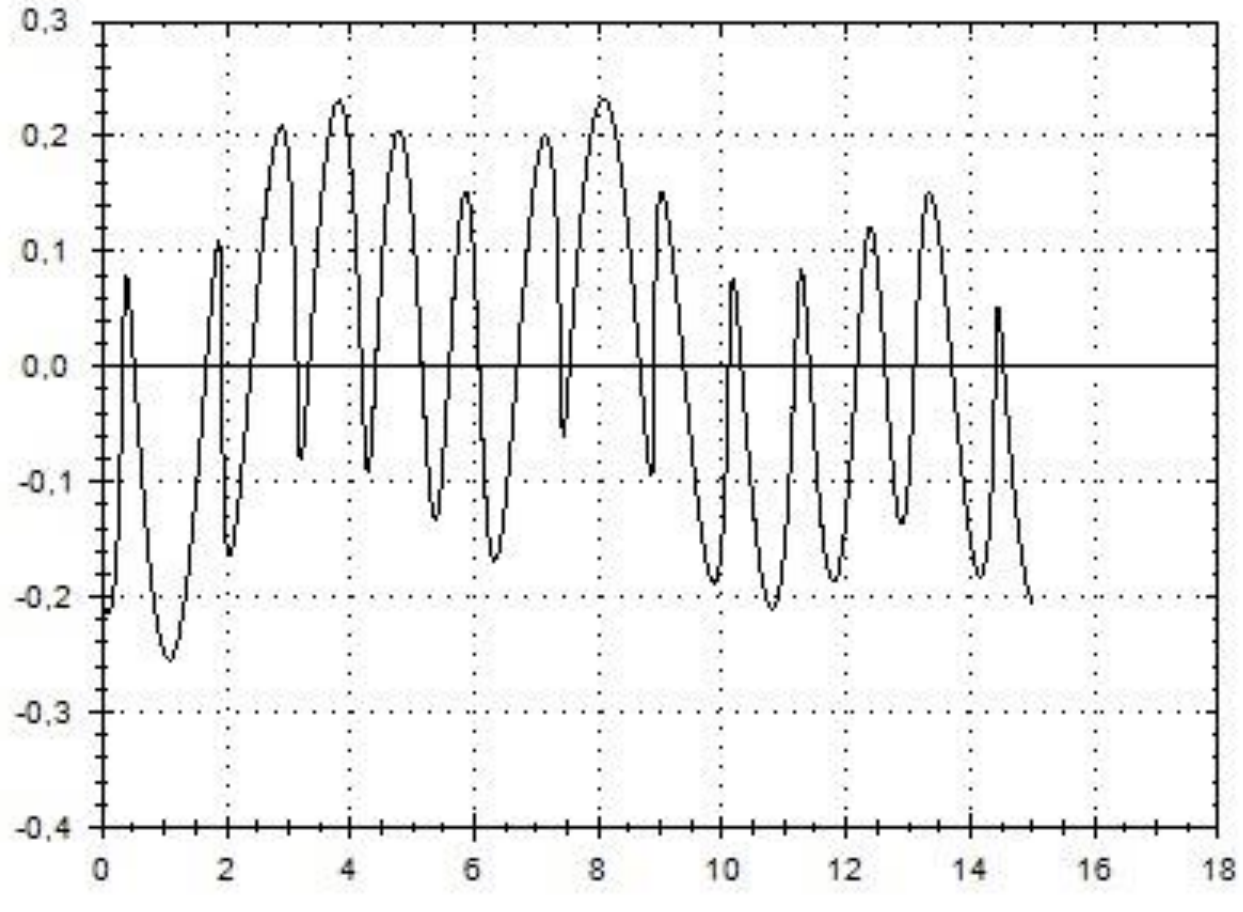

Fig.7. Numerical solution for $(\bar{z})$.

Thus, trajectories have the stable dynamics (for the chosen initial conditions, including those $\{\bar{z}_0 = -0.2, (\dot{\bar{z}})_0 = -0.3\}$ for coordinate $(\bar{z})$) without sudden jumping of the solutions.

## Acknowledgment.

Authors appreciate participation of Dr. Tetiana Kozachenko in numerical experiments at earlier stages of this work.

On behalf of all authors, the corresponding author should confirm that there is no conflict of interest regarding this work. The data for this paper are available by contacting the corresponding author.

In this research, all the authors agreed with results and conclusions of each other in Sections 1-6.